# Roadmap for GaAs spin qubits

Ferdinand Kuemmeth[1] and Hendrik Bluhm[2]

**Status**

Gate-defined quantum dots in GaAs have been used extensively for pioneering spin qubit devices due to the relative simplicity of fabrication and favourable electronic properties such as a single conduction band valley, a small effective mass, and stable dopants. Decades of prior improvements of the growth of III-V heterostructures by molecular beam epitaxy had resulted in the availability of high-quality substrates for various applications, and spin qubits were ultimately first demonstrated in GaAs in 2005, significantly before the first Si qubits in 2012. GaAs spin qubits are now readily produced in many labs, whereas the realization of comparable devices in Si remains challenging. However, a disadvantage is the unavoidable presence of nuclear spins, leading to an intrinsic $T_2^*$ of about 10 ns. Dynamical decoupling can extend the coherence time to the millisecond range [1], and single-qubit control with a fidelity of 99.5 % was demonstrated [2]. Nevertheless, these techniques require a significant effort in controlling and suppressing nuclear spin fluctuations, and so far have only been successful for singlet-triplet qubits encoded in two-electron spin states associated with double quantum dots. GaAs quantum dots are currently used as a testbed for entanglement [3], quantum non-demolition measurements [4], automatic tuning [5, 6], multi-dot arrays [7, 8], coherent exchange coupling [8], teleportation [9] etc., partly because reproducible Si devices are not broadly available yet. Much of the resulting insights can be transferred to Group IV material systems, although specific properties of GaAs are also actively studied. Remarkable recent achievements include the transfer of electrons between quantum dots using surface acoustic waves [10], which could be used to overcome the challenge of connecting distant qubits, and the detection of photo-generated carriers, a precursor to the ability to convert flying photonic qubits into spin states [11]. Last but not least, qubits in GaAs quantum dots are of interest as a manifestation of quantum many-body physics, such as the central spin problem or itinerant magnetism [12].

**Current and Future Challenges**

The operation of gate-defined spin qubits relies on voltages – quasistatic voltages for tuning the device to an appropriate operating point, and time-dependent control voltages for the coherent manipulation on nanosecond timescales – which in a modern dilution refrigerator should be practical up to approximately 100 qubits. On the flipside, this makes the quantum processor susceptible to effective electrical noise, requiring a careful trade-off of instrumentation noise and the material's intrinsic charge noise against other engineering constraints. Just like the encoding in specific two-electron spin states makes a singlet-triplet qubit robust to global magnetic field fluctuations, other encodings in three-electron [13] or four-electron [14] spin states have recently been proposed that also mitigate noise in the magnetic gradient between dots (particularly relevant for GaAs) and effective charge noise (relevant also for Si). The role of symmetric operating points in these proposals are being experimentally studied in GaAs multi-dot arrays [15], exposing a new engineering challenge: The large number of physical gate electrodes per quantum dot (facilitated by the relatively large size of GaAs quantum dots) allows independent tuning of many local degrees of freedom (dot occupation, interdot tunnel barriers, etc.), but ultimately will impose unrealistic wiring requirements.

[1] *Center for Quantum Devices, Niels Bohr Institute, University of Copenhagen, Denmark*
[2] *JARA-FIT Institute for Quantum Information, RWTH Aachen University and Forschungszentrum Jülich, Germany*

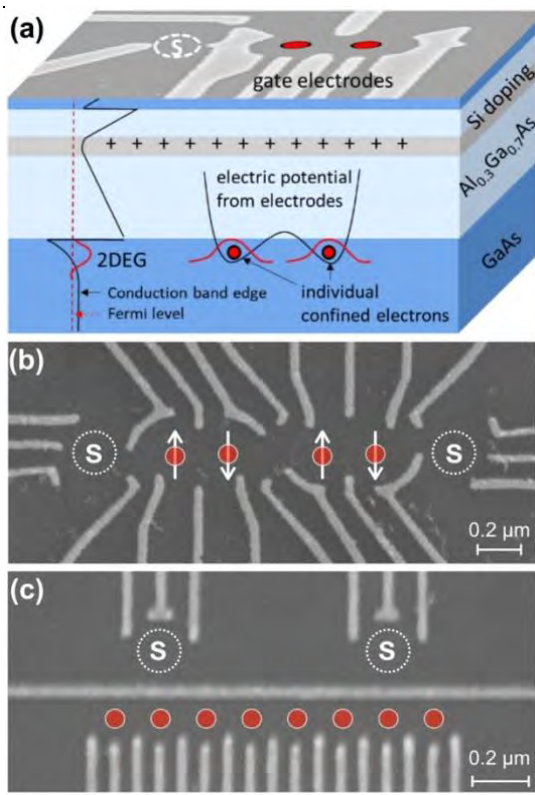

Figure 1. Representative GaAs qubit devices – from single and double dots to linear arrays.
(a) Top-gated GaAs heterostructure resulting in controllable one-electron quantum dots with proximal charge sensor (S) for readout. (b) Two proximal double dots to study entanglement between two nearest-neighbour singlet-triplet qubits [3]. (c) Progress towards linear spin chains [6]. Figure credits: Hendrik Bluhm, RWTH Aachen (a), Shannon Harvey, group of Amir Yacoby at Harvard University (b), Christian Volk, group of Lieven Vandersypen, TU Delft (c).

For a processor with more than 1000 spin qubits, a radical change will be needed on how to integrate quantum dots at cryogenic temperatures with scalable control electronics. Even for current devices, the ultimate limits of coherence and control fidelity are still uncharted, despite the fact that the nature of the hyperfine coupling between electron and nuclear spins is rather well known and many of the resulting effects are now understood in considerable detail. Using appropriate control pulse optimization, substantial improvements in the demonstrated two-qubit gate fidelities can be expected. As for all types of quantum dot qubits, a mechanism for high-fidelity long-range coupling would likely be required for truly scalable quantum circuits, potentially building upon current efforts to couple GaAs dots to superconducting cavities [16] [cf. Fig. 2(d)] or shuttling of electrons [Fig. 2(e)]. Although anecdotal experience in many labs points to a good reproducibility of GaAs quantum dots, no systematic study supports this evidence, and the limiting factors are unknown. A detailed yield investigation could reveal if the small effective mass is a decisive advantage and could serve as a reference benchmark for Si-based devices.

**Advances in Science and Technology to Meet Challenges**
Further improvement of coherence and control fidelity will benefit from both improved dynamic nuclear polarization procedures to suppress fluctuations of the hyperfine field as well as a reduction of charge noise. Somewhat surprisingly, simulations indicate that charge noise is the more limiting factor. For long range coupling approaches via cavities or electron shuttling, material-specific limitations will have to be understood. Piezoelectricity, spin-orbit coupling, and nuclear spins work against the GaAs material system, whereas the single valley and small mass are advantages. For cavity coupling, current performance metrics are not nearly good enough for high-fidelity entangling gates. From the quantum control point of view, one challenge appears to be that optimal pulses require careful cancelation of errors due to quasi-static noise. Applying simulated pulses in experiments may compromise the desired performance due to imperfect system knowledge and thus require new approaches to gate characterization and calibration.

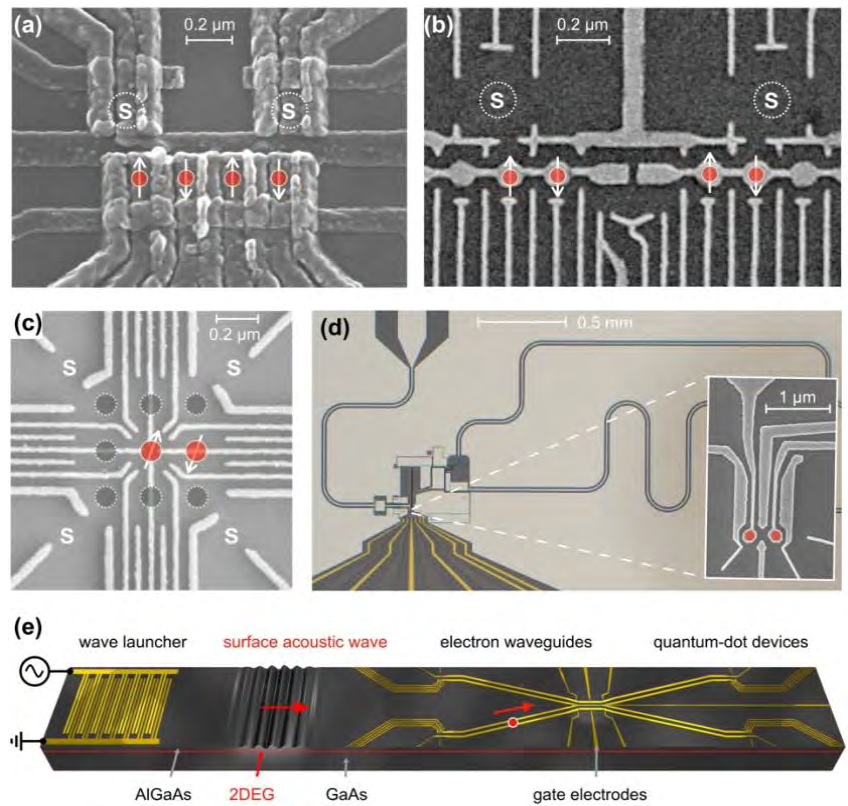

**Figure 2**. **Current approaches to larger quantum circuits with GaAs qubits**.   Intermediate distance coupling by **(a)** Heisenberg teleportation [9] and **(b)** mediated exchange [8]. First steps towards larger circuits by **(c)** operating spins within a small 2D array [7], **(d)** integration with superconducting devices and circuit quantum electrodynamics [16], and by **(e)** moving electrons via surface acoustic waves [10]. Figure credits: John Nichol, University of Rochester (a), Center for Quantum Devices, University of Copenhagen (b), Tristan Meunier, Université Grenoble Alpes (c), Pasquale Scarlino, Group of Andreas Wallraff, ETH Zurich (d), Christopher Bäuerle, Université Grenoble Alpes (e).

Regarding device designs, more complicated circuits would greatly benefit from multiple metal layers, as shown in Fig. 2(a). Yet, the then required dielectrics may be an additional source of charge noise and will sacrifice some of the fabrication simplicity. An exciting prospect associated with the direct band gap is to convert between spin and photon states, or to entangle them. This capability could be a major advantage over Si by allowing the realization of networks of quantum processors for communication and distributed computing and by opening additional options for long-range on-chip coupling. Much of the fundamental principles have been demonstrated using self-assembled quantum dots and could be transferred to hybrid devices with some kind of exciton trap coupled to a gate-defined dot [17]. However, such devices yet remain to be realized.

**Concluding Remarks**
GaAs-based devices have been crucial for the birth of quantum dot qubits. Much attention is now shifting to Si. While the reasons for this trend are largely compelling, it is not established with scientific rigour that Si is preferable to GaAs when considering all factors. The compatibility of Si with CMOS processing is often seen as an advantage. However, one should also keep in mind that process development for the unusual layouts compared to transistors with small feature sizes needed for Si qubits will incur large development costs for foundry fabrication. In any case, GaAs devices will likely remain a workhorse for proof-of-concept quantum information processing and solid-state experiments. Considerable technological and scientific potential may arise from advances in optical coupling.

## References

[1] *Notch filtering the nuclear environment of a spin qubit*, F. K. Malinowski, F. Martins, P. D. Nissen, E. Barnes, M. S. Rudner, S. Fallahi, G. C. Gardner, M. J. Manfra, C. M. Marcus, F. Kuemmeth, Nature Nanotechnology **12**, 16 (2016).

[2] *Closed-loop control of a GaAs-based singlet-triplet spin qubit with 99.5% gate fidelity and low leakage*, P. Cerfontaine, T. Botzem, J. Ritzmann, S. S. Humpohl, A. Ludwig, D. Schuh, D. Bougeard, A. D. Wieck, H. Bluhm, arXiv:1906.06169 (2019).

[3] *High-fidelity entangling gate for double-quantum-dot spin qubits*, J. M. Nichol, L. A. Orona, S. P. Harvey, S. Fallahi, G. C. Gardner, M. J. Manfra & A. Yacoby, npj Quantum Information **3,** 3 (2017).

[4] *Quantum non-demolition measurement of an electron spin qubit*, T. Nakajima, A. Noiri, J. Yoneda, M. R. Delbecq, P. Stano, T. Otsuka, K. Takeda, S. Amaha, G. Allison, K. Kawasaki, A. Ludwig, A. D. Wieck, D. Loss & S. Tarucha, Nature Nanotechnol. **14**, 555 (2019).

[5] *A machine learning approach for automated fine-tuning of semiconductor spin qubits*, J. D. Teske, S. S. Humpohl, R. Otten, P. Bethke, P. Cerfontaine, J. Dedden, A. Ludwig, A. D. Wieck, H. Bluhm, Appl. Phys. Lett. **114**, 133102 (2019).

[6] *Loading a quantum-dot based "Qubyte" register*, C. Volk, A. M. J. Zwerver, U. Mukhopadhyay, P. T. Eendebak, C. J. van Diepen, J. P. Dehollain, T. Hensgens, T. Fujita, C. Reichl, W. Wegscheider, L. M. K. Vandersypen, npj Quantum Information **5**, 29 (2019).

[7] *Coherent control of individual electron spins in a two dimensional array of quantum dots*, P.-A. Mortemousque, E. Chanrion, B. Jadot, H. Flentje, A. Ludwig, A. D. Wieck, M. Urdampilleta, C. Bauerle, T. Meunier, arXiv:1808.06180 (2018).

[8] *Fast spin exchange across a multielectron mediator*, F. K. Malinowski, F. Martins, T. B. Smith, S. D. Bartlett, A. C. Doherty, P. D. Nissen, S. Fallahi, G. C. Gardner, M. J. Manfra, C. M. Marcus & F. Kuemmeth, Nature Communications **10**, 1196 (2019).

[9] *A Heisenberg Spin Teleport*, Y. P. Kandel, H. Qiao, S. Fallahi, G. C. Gardner, M. J. Manfra, J. M. Nichol, Nature **573**, 553 (2019).

[10] *Sound-driven single-electron transfer in a circuit of coupled quantum rails*, S. Takada, H. Edlbauer, H. V. Lepage, J. Wang, P.-A. Mortemousque, G. Georgiou, C. H. W. Barnes, C. J. B. Ford, M. Yuan, P. V. Santos, X. Waintal, A. Ludwig, A. D. Wieck, M. Urdampilleta, T. Meunier, C. Bäuerle, Nature Communications **10**, 4557 (2019).

[11] *Nondestructive Real-Time Measurement of Charge and Spin Dynamics of Photoelectrons in a Double Quantum Dot*, T. Fujita, H. Kiyama, K. Morimoto, S. Teraoka, G. Allison, A. Ludwig, A. D. Wieck, A. Oiwa, and S. Tarucha, Phys. Rev. Lett. **110**, 266803 (2013).

[12] *Nagaoka ferromagnetism observed in a quantum dot plaquette*, J. P. Dehollain, U. Mukhopadhyay, V. P. Michal, Y. Wang, B. Wunsch, C. Reichl, W. Wegscheider, M. S. Rudner, E. Demler, L. M. K. Vandersypen, arXiv:1904.05680 (2019).

[13] *Exchange-only singlet-only spin qubit*, A. Sala and J. Danon, Phys. Rev. B **95**, 241303(R) (2017).

[14] *Quadrupolar Exchange-Only Spin Qubit*, M. Russ, J. R. Petta, and G. Burkard, Phys. Rev. Lett. **121**, 177701 (2018).

[15] *Symmetric operation of the resonant exchange qubit*, F. K. Malinowski, F. Martins, P. D. Nissen, S. Fallahi, G. C. Gardner, M. J. Manfra, C. M. Marcus, and F. Kuemmeth, Phys. Rev. B **96**, 045443 (2017).

[16] *Coherent microwave-photon-mediated coupling between a semiconductor and a superconducting qubit*, P. Scarlino, D. J. van Woerkom, U. C. Mendes, J. V. Koski, A. J. Landig, C. K. Andersen, S. Gasparinetti, C. Reichl, W. Wegscheider, K. Ensslin, T. Ihn, A. Blais and A. Wallraff, Nature Communications **10**, 3011 (2019).

[17] *Optically Loaded Semiconductor Quantum Memory Register*, D. Kim, A. A. Kiselev, R. S. Ross, M. T. Rakher, C. Jones, and T. D. Ladd, Phys. Rev. Applied **5**, 024014 (2016).